\documentclass[12pt,a4paper,tightenlines]{revtex4}
\usepackage{amssymb,amsmath}
\usepackage{psfrag}
\usepackage[dvips]{graphicx}

\begin{document}

\title{Low pumping energy mode of the ``optical bars''/``optical lever''
topologies of gravitational-wave antennae}

\author{F.Ya.Khalili}

\affiliation{Physics Faculty, Moscow State University, Moscow, Russia}

\email{farid@hbar.phys.msu.su}


\begin{abstract}

The ``optical bars''/``optical lever'' topologies of gravitational-wave
antennae allow to obtain sensitivity better that the Standard Quantum Limit
while keeping the optical pumping energy in the antenna relatively low.
Element of the crucial importance in these schemes is the local meter which
monitors the local test mirror position. Using cross-correlation of this meter
back-action noise and its measurement noise it is possible to further decrease
the optical pumping energy. In this case the pumping energy minimal value will
be limited by the internal losses in the antenna only. Estimates show that for
values of parameters available for contemporary and planned gravitational-wave
antennae, sensitivity about one order of magnitude better than the Standard
Quantum Limit can be obtained using the pumping energy about one order of
magnitude smaller energy than is required in the traditional topology in order
to obtain the the Standard Quantum Limit level of sensitivity.

\end{abstract}

\maketitle

\section{Introduction}

First generation large-scale laser interferometric gravitational-wave antennae
\cite{300years, Abramovici1992} are being placed into operation nowadays
\cite{LIGOSite}. The second generation of laser gravitational-wave antennae
development is under way concurrently \cite{WhitePaper1999}. Sensitivity of
these second generation antennae will be close to the Standard Quantum Limit
(SQL) \cite{67a1eBr} that is characteristic sensitivity level where the
measurement noise ({\it i.e.} the shot noise in the laser interferometric
devices) and the back-action noise ({\it i.e.} the radiation pressure noise)
contribute equal parts to the measurement error:

\begin{equation}\label{SQL}
  S_h^{\rm SQL}(\Omega) = \frac{4\hbar}{M\Omega^2L^2} \,.
\end{equation}
Here $S_h^{\rm SQL}(\Omega)$ is the double-sided spectral density of the
equivalent noise for the dimensionless amplitude of the metrics perturbation
$h(t)$, $\Omega$ is the mean frequency of the gravitational-wave signal, $M$
is the mass of the interferometer mirrors, and $L$ is the length of the
interferometer arms.

Further improvement of the sensitivity will require to use the Quantum
Non-Demolition (QND) measurement methods \cite{77a1eBrKhVo, Thorne1978,
80a1BrVoTh} which allow to eliminate the part produced by the back-action
noise from the meter output. Several possible design of QND laser
gravitational-wave antennae have been proposed already. They can be divided
into three main groups.

The first group is based on the fact that the value of the SQL depends on the
nature of the test object. In particular, a harmonic oscillator provides
better sensitivity in the narrow band close to its resonance frequency than a
free mass does, even if the same SQL-limited meter is used in both cases
\cite{99a1BrKh}. It was shown in articles \cite{Buonanno2001, Buonanno2002}
that in the signal-recycling configuration of the interferometric
gravitational-wave antennae an optical rigidity can be created rather easily
that will turn test masses into mechanical oscillators. Moreover, this optical
rigidity has specific spectral dependence which allows to obtain sensitivity
better than the SQL for both free mass and ordinary harmonic oscillator. Using
this method it is possible to obtain sensitivity a few times better than the
SQL for a free mass in a relatively wide band \cite{Buonanno2001,
Buonanno2002} or ``dive'' deep below the SQL in a narrow band \cite{01a2Kh}.
It is necessary to note that both regimes require about the same optical
pumping energy as standard SQL-limited schemes. It is the author's opinion
that these relatively simple methods could (and should) be implemented already
in the second generation of laser gravitational-wave antennae.

The second group of methods requires more substantial modifications of the
laser gravitational-wave antenna topology which convert it into a QND device.
Examples of this approach are: interferometer with modified input and/or
output optics, which implements the spectral variational measurement
\cite{02a1KiLeMaThVy} and different implementations of the {\em quantum
speedmeter} scheme \cite{00a1BrGoKhTh, Purdue2001, Purdue2002, Chen2002,
02a2Kh}. In principle, they allow to obtain arbitrarily high sensitivity. In
practice, however, they ``suffer'' from very high optical power circulating in
the interferometer arms, which also depends sharply on the required
sensitivity.

It can be shown (see \cite{00p1BrGoKhTh}) that the optical energy has not to
be smaller than the value of

\begin{equation}\label{E_EQL}
  \mathcal{E} = \frac{{\cal E}_{\rm SQL}\zeta^2}{2\xi^2}\,
\end{equation}
where $\xi$ is the ratio of the signal amplitude which can be detected to the
amplitude corresponding to the Standard Quantum Limit (the smaller is $\xi$
the better is the sensitivity), and $\zeta$ is the squeezing factor ($\zeta=1$
for the optical field coherent quantum state and $\zeta<1$ for the squeezed
state),

\begin{equation}\label{E_SQL}
  {\cal E}_{\rm SQL} = \frac{ML^2\Omega^3}{2\omega_o}
\end{equation}
is the optimal energy for the SQL-limited interferometric antenna, $\omega_o$
is the pumping frequency, and $L$ is the length of the interferometer arms
\footnote{Numerical factors in formulae (\ref{SQL}, \ref{E_EQL}, \ref{E_SQL})
correspond to the case of the standard Michelson---Fabry-Perot topology with
four mirrors having equal masses $M$, which was analyzed in detail in article
\cite{02a1KiLeMaThVy}. Note also the factor $1/2$ in formula (\ref{E_EQL}),
which arises for the back-action noise that makes up half of the net noise in
the SQL-limited detectors, is eliminated from the output signal of the QND
detectors. It allows to obtain sensitivity equal to the SQL ($\xi=1$) using
two times smaller energy or, put it otherwise, to obtain sensitivity $\sqrt2$
times better than the SQL ($\xi=1/\sqrt2$) using ${\cal E}={\cal E}_{\rm
SQL}$.}. Formula (\ref{E_SQL}) is valid in the wide-band regime where the
bandwidth $\Delta\Omega\sim\Omega$. In the narrow-band regime the energy
can be reduced by the factor of $\sim\Delta\Omega/\Omega$.

If, for example, $M=40\,{\rm Kg}$, $L=4\,{\rm Km}$, $\Omega=2\pi\times
100\,{\rm s}^{-1}$ and $\omega_o=2\times 10^{15}\,{\rm s}^{-1}$ (these values
correspond to the proposed gravitation-wave antenna LIGO-II
\cite{WhitePaper1999}) then ${\cal E}_{\rm SQL}\approx 40\,{\rm J}$.
Corresponding circulating optical pumping power is equal to

\begin{equation}
  W_{\rm SQL} = \frac{c}{4L}\,{\cal E}_{\rm SQL} \approx 0.75\,{\rm MWt} \,.
\end{equation}
It is possible to conclude that the feasibility of these second group methods
depends crucially on the experimental progress in preparation of highly
squeezed quantum states (with $\zeta\ll 1$).

The third and the most radical approach, {\em intracavity readout scheme}, was
proposed in the article \cite{96a2BrKh}. It was proposed to measure directly
the redistribution of the optical energy {\em inside} the antenna in a QND way
(without absorption of the optical quanta) instead of monitoring output light
beam {\em outside} the antenna using photodetectors. In this case the
necessary non-classical quantum state of the optical field [factor $\zeta$ in
the formula (\ref{E_EQL})] is generated automatically and therefore the
pumping energy does not depend directly on the required sensitivity.

\begin{figure}[t]

\psfrag{A}{{\sf A}}
\psfrag{B}{{\sf B}}
\psfrag{C}{{\sf C}}

\includegraphics[width=4in]{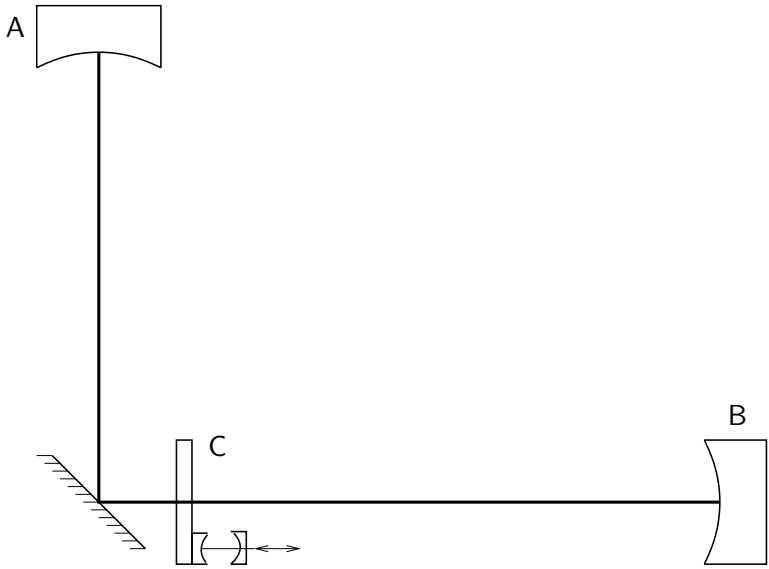}

\caption{The ``optical bars'' intracavity scheme}\label{fig:optbar}

\end{figure}

In the article \cite{97a1BrGoKh} a possible implementation of the
gravitational-wave antenna with intracavity detection, so-called ``optical
bars'' scheme, was proposed (see FIG.\,\ref{fig:optbar}). In this scheme end
mirrors {\sf A}, {\sf B}, and an additional local mirror {\sf C} form two
Fabry-Perot cavities coupled by means of the mirror {\sf C} amplitude
transmittance $T$. Relatively weak external pumping (not shown in the picture)
is necessary in order to compensate internal losses in the optical elements
and to support steady value of the optical energy in the cavities. The optical
field acts here as a two rigid springs with one located between the mirrors
{\sf A} and {\sf C}, while the second one (L-shaped) located between the
mirrors {\sf B} and {\sf C}. The rigidity of these springs is equal to

\begin{equation}
  K = \frac{2\omega_o{\cal E}}{L^2\Omega_B} \,,
\end{equation}
where

\begin{equation}
  \Omega_B = \frac{cT}{L} \,.
\end{equation}
is the sloshing frequency of the system of two coupled cavities {\sf AC} and
{\sf BC} [strictly speaking, $K$ is equal to the asymptotic value of the
rigidity at $\Omega\ll\Omega_B$; see formula (\ref{cal_K})].

Due to these springs displacement of the end mirrors {\sf A} and {\sf B}
caused by the gravitational wave produces displacement of the local mirror
{\sf C}, which can be detected, for example, by a small-scale optical
interferometric meter which monitors position of the mirror {\sf C} relative
to reference mass placed outside the optical pumping field.

It was shown in the article \cite{97a1BrGoKh} that if optical energy exceeds
some threshold value of ${\cal E}_{\rm thres}$ then these springs are rigid
enough to provide the signal displacement of the local mirror {\sf C} equal to
the displacement of the end mirrors. In this case the sensitivity does not
depend on the optical energy and is defined by the sensitivity of the local
meter only (compare with formula (\ref{E_EQL}) which contains the factor
$\xi^{-2}$). However, it was concluded in the article \cite{97a1BrGoKh} that
the threshold energy have to be rather large and close to ${\cal E}_{\rm
SQL}$.

\begin{figure}[t]

\psfrag{A}{{\sf A}}
\psfrag{B}{{\sf B}}
\psfrag{A1}{{\sf A'}}
\psfrag{B1}{{\sf B'}}
\psfrag{C}{{\sf C}}

{\includegraphics[width=4in]{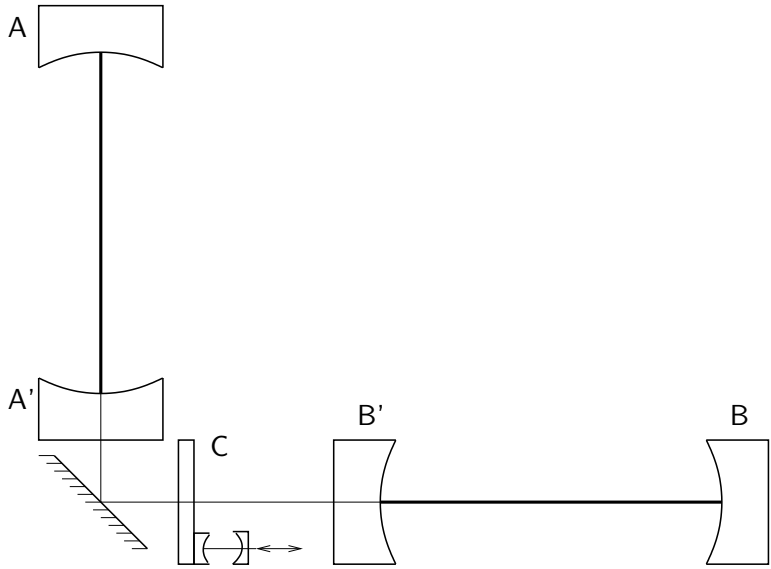}}

\caption{The ``optical lever'' intracavity scheme}\label{fig:optlever}

\end{figure}

In the article \cite{02a1Kh} an improved version of the ``optical bars''
scheme was considered. It differs from the original ``optical bar'' scheme by
two additional mirrors {\sf A'} and {\sf B'} (see FIG.\,\ref{fig:optlever})
which turn the antenna arms into two Fabry-Perot cavities {\sf AA'} and {\sf
BB'} similar to the standard Fabry-Perot --- Michelson topology of the
contemporary gravitational-wave antennae. This scheme was called the ``optical
lever'' because it can provide significant gain in the signal displacement of
the local mirror similar to the gain which can be obtained using ordinary
mechanical lever with unequal arms. The value of this gain is equal to

\begin{equation}
  \digamma = \frac{2}{\pi}\,{\cal F} \,,
\end{equation}
where ${\cal F}$ is the finesse of the Fabry-Perot cavities {\sf AA'} and {\sf
BB'}. It was shown in the article \cite{02a1Kh} that in all other aspects the
``optical lever'' scheme is identical to the ``optical bars'' one but in the
former one the local mirror ${\sf C}$ transmittance $T$ have to be $\digamma$
times larger, and its mass have to be $\digamma^2$ times smaller. Due to this
scaling of mass the gain in the signal displacement by itself does not allow
to overcome the SQL as the SQL value increases exactly in the same proportion.
But it allows to use less sensitive local position meter and increases the
signal-to-noise ratio for miscellaneous noises of non-quantum origin.

No means of reducing the optical power below the level of ${\cal E}_{\rm SQL}$
were discussed in the article \cite{02a1Kh}.

At the same time different regimes of the ``optical bars'' scheme were
considered in brief in the article \cite{98a1BrGoKh}, and it was mentioned
that if a local meter with cross-correlated measurement noise and back-action
noise is used then the threshold energy can be substantially lower than the
${\cal E}_{\rm SQL}$ (see subsection 4.2 of that article). It is evident that
this conclusion is also valid for the ``optical lever'' scheme.

In the current article this cross-correlation regime is analyzed in detail. In
the section \ref{sec:mechmodel} the simple mechanical model is considered,
which shows how the cross-correlation of the meter noises allows to reduce the
pumping energy. In the section \ref{sec:sensitivity} sensitivity limits are
considered and numerical estimates are provided. Most of the calculations are
done for the more simple ``optical bars'' scheme; applicability for the
``optical lever'' version is considered in the subsection \ref{sec:optlever}.

\section{Mechanical model}\label{sec:mechmodel}

\begin{figure}[t]

\begin{center}

\psfrag{M}{$M$}
\psfrag{m}{$m$}
\psfrag{x1}{$x_1$}
\psfrag{x2}{$x_2$}
\psfrag{y}{$y$}
\psfrag{F}[cb]{$\frac12F_{\rm grav}$}
\psfrag{K}{$\frac12K$}

\includegraphics[width=4in]{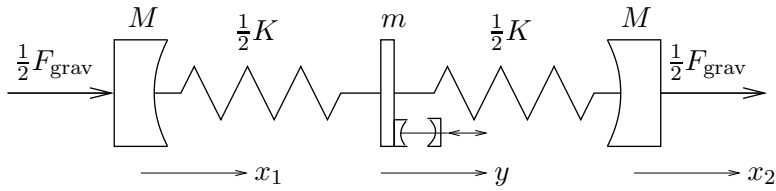}

\end{center}

\caption{Mechanical model}\label{fig:mechmodel}

\end{figure}

Consider the mechanical model shown in FIG.\,\ref{fig:mechmodel}. Here masses
$M$ correspond to the end mirrors of the gravitational-wave antenna (compare
with FIG.\,\ref{fig:optbar}), mass $m$ corresponds to the local mirror, and
springs $K/2$ correspond to the optical rigidity. Equal signal forces

\begin{equation}\label{F_grav}
  \frac{F_{\rm grav}}{2}=\frac{M\ddot h}{2}
\end{equation}
acts on both end masses $M$, and the goal is to detect the signal by
monitoring the local mass $m$ state.

Suppose that conditions for the ``intermediate case'' of the article
\cite{98a1BrGoKh} are fulfilled, namely, the local mass $m$ is small,

\begin{equation}\label{small_m}
  m\Omega^2 \ll K \,,
\end{equation}
where $\Omega$ is the signal frequency and the rigidity $K$ is also relatively
small:

\begin{equation}\label{smallK}
  K \ll 2M\Omega^2 \,,
\end{equation}
Due to condition (\ref{small_m}) the signal displacement of the local mass
will be equal to the signal displacement of the end ones. However, due to
condition (\ref{smallK}) the Standard Quantum Limit for the position of the
local test mass $\Delta y_{\rm SQL}$ will be much larger than the Standard
Quantum Limit for the positions of the heavy end masses

\begin{equation}\label{SQL_x}
  \Delta x_{\rm SQL} \simeq \sqrt{\frac{\hbar}{M\Omega^2\tau}} \,.
\end{equation}
Really, let $\Delta y_{\rm meas}$ be the measurement precision provided by the
local meter.  Due to the uncertainty relation perturbation of the mass $m$
momentum in this case will be equal to

\begin{equation}
  \Delta p_{\rm pert} = \frac{\hbar}{2\Delta y_{\rm meas}}\,.
\end{equation}
It corresponds to the random force with the uncertainty equal to

\begin{equation}
  \Delta F_{\rm pert} = \frac{\hbar}{2\Delta y_{\rm meas}\tau}\,,
\end{equation}
which produces the additional random displacement of the local test mass

\begin{equation}
  \Delta y_{\rm pert} = \frac{\hbar}{2K\Delta y_{\rm meas}\tau}\,.
\end{equation}
Therefore, the sum error will be equal to

\begin{equation}
  \Delta y = \sqrt{(\Delta y_{\rm meas})^2
    + \left(\frac{\hbar}{K\Delta y_{\rm meas}\tau}\right)^2}.
 \end{equation}
Minimum of this expression is equal to

\begin{equation}
  \Delta y_{\rm SQL} = \sqrt{\frac{\hbar}{2K\tau}} \,,
\end{equation}
and this value is $M\Omega^2/K\gg 1$ times larger than the Standard Quantum
Limit (\ref{SQL_x}). Due to this consideration it was concluded in the article
\cite{97a1BrGoKh} that in order to obtain sensitivity close to the $\Delta
x_{\rm SQL}$ it is necessary to use strong rigidity:

\begin{equation}
  K \gtrsim M\Omega^2 \,,
\end{equation}
and therefore, large pumping energy.

Consider, however, the situation more precisely. If condition (\ref{small_m})
is fulfilled then equations of motion (in the spectral representation) of the
system shown in FIG.\,\ref{fig:mechmodel} looks like:

\begin{align}
  -2M\Omega^2\hat x(\Omega) + K\hat x(\Omega)
    &= K\hat y(\Omega) + F_{\rm grav}(\Omega) \,, \\
  K\hat y(\Omega) &= K\hat x(\Omega) + \hat F_{\rm meter}(\Omega) \,,
\end{align}
where $x=(x_1+x_2)/2$ and $\hat F_{\rm meter}$ is the back-action force of the
local meter.

It follows from these equations that the output signal of the meter is equal
to:

\begin{equation}\label{tilde_y_1}
  \tilde y(\Omega)
  = \frac{F_{\rm grav}(\Omega) + \hat F_{\rm meter}(\Omega)}{-2M\Omega^2}
    + \frac{\hat F_{\rm meter}(\Omega)}{K} + \hat y_{\rm
    meter}(\Omega)\,,
\end{equation}
where $\hat y_{\rm meter}$ is the measurement noise which determines the
measurement error $\Delta y_{\rm meas}$.

It is easy to see that if the measurement noise contains the part
proportional to the back-action force:

\begin{equation}\label{correl_simple}
  \hat y_{\rm meter} = y_{\rm meter}^{(0)} - \frac{F_{\rm meter}}{K} \,,
\end{equation}
then the main back-action term $\hat F_{\rm meter}/K$ in the equation
(\ref{tilde_y_1}) vanishes:

\begin{equation}\label{correl_out_1}
  \tilde y(\Omega)
  = \frac{F_{\rm grav}(\Omega) + \hat F_{\rm meter}(\Omega)}{-2M\Omega^2}
    + \hat y_{\rm meter}^{(0)}(\Omega) \,.
\end{equation}
Of course strong perturbation $\hat F_{\rm meter}/K$ still exists in this case
but the meter does not ``see'' it as it is masked by the correlated with
$F_{\rm meter}$ part of $y_{\rm meter}$.

Equation (\ref{correl_out_1}) corresponds exactly to the output signal of
the position meter with the measurement noise $\hat y_{\rm meter}^{(0)}$
attached directly to a test mass $2M$. In particular, if an ordinary
interferometric position meter with constant pumping power and time- and
frequency-independent phase is used, then sensitivity of such a scheme will be
limited by the SQL (\ref{SQL_x}) even if the rigidities are much smaller than
$M\Omega^2$, as long as condition (\ref{small_m}) is fulfilled.

\section{Sensitivity Limits}\label{sec:sensitivity}

We will consider here only two main factors which limit the sensitivity of the
scheme being considered: optical losses and noises of the local meter putting
aside numerous sensitivity limits which are common for all topologies of the
laser gravitational-wave antennae (in particular, miscellaneous internal
noises in the mirrors and mirror suspensions). In this case output signal of
this scheme normalized as an equivalent gravitational-wave signal can be
presented as

\begin{equation}
  \tilde h = h + h_{\rm loss} + h_{\rm meter} \,,
\end{equation}
where $h$ is the actual gravitational-wave signal, $h_{\rm loss}$ is the noise
which arises due to the optical losses, and $h_{\rm meter}$ is the one created
by the local meter fluctuations. These noises are calculated in appendix
\ref{app:noises}.

\subsection{Optical losses}\label{opt_losses}

Taking into account formulae (\ref{h_loss}), (\ref{S_loss_raw}), and
approximations which have been made in the appendix \ref{app:noises}, spectral
density of the noise $h_{\rm loss}$ can be presented as

\begin{equation}\label{S_loss}
  S_h^{\rm loss} = \frac{2\hbar L^2\gamma}{\omega_o{\cal E}} \,,
\end{equation}
where $\gamma$ is the Fabry-Perot cavities damping rate. Ratio of this
spectral density to the spectral density (\ref{SQL}) which corresponds to the
SQL is equal to \footnote{The last term in formula (\ref{xi2_loss}) looks a
bit misleading because it seems that the smaller is $\Omega$ the worse is
sensitivity. It is easy to see that it is not the case because ${\cal E}_{\rm
SQL}$ also depends on $\Omega$ and the worst case takes place at the highest
signal frequency. Therefore, it is for this frequency all estimates should be
done.}

\begin{equation}\label{xi2_loss}
  \xi_{\rm loss}^2 = \frac{S_h^{\rm loss}}{S_h^{\rm SQL}}
  = \frac{M\Omega^2\gamma}{2\omega_o{\cal E}}
  = \frac{{\cal E}_{\rm SQL}}{{\cal E}}\,\frac{\gamma}{\Omega} \,.
\end{equation}
This formula resembles formula (\ref{E_SQL}) because in both these formulae
the larger is ratio ${\cal E}/{\cal E}_{\rm SQL}$ the better is the
sensitivity. At the same time formula (\ref{xi2_loss}) contains the factor
$\gamma/\Omega$ which in principle can be made very small. In particular,
modern achievements in fabrication of high-quality mirrors permits to obtain
$\gamma\lesssim 1\,{\rm s}^{-1}\simeq 10^{-3}\,\Omega$. Therefore, even, say,
for ${\cal E}\simeq 0.1{\cal E}_{\rm SQL}$ sensitivity of about one order of
magnitude better than the SQL, $\xi_{\rm loss}\simeq 0.1$ could be achieved.

\subsection{SQL-limited local meter}

\begin{figure}[t]

\begin{center}

\psfrag{x}{$\Omega/\Omega_{\rm SQL}$}
\psfrag{y}
  {$\dfrac{S_h^{\rm meter}(\Omega)}{S_h^{\rm asympt}(\Omega_{\rm SQL})}$}

\includegraphics{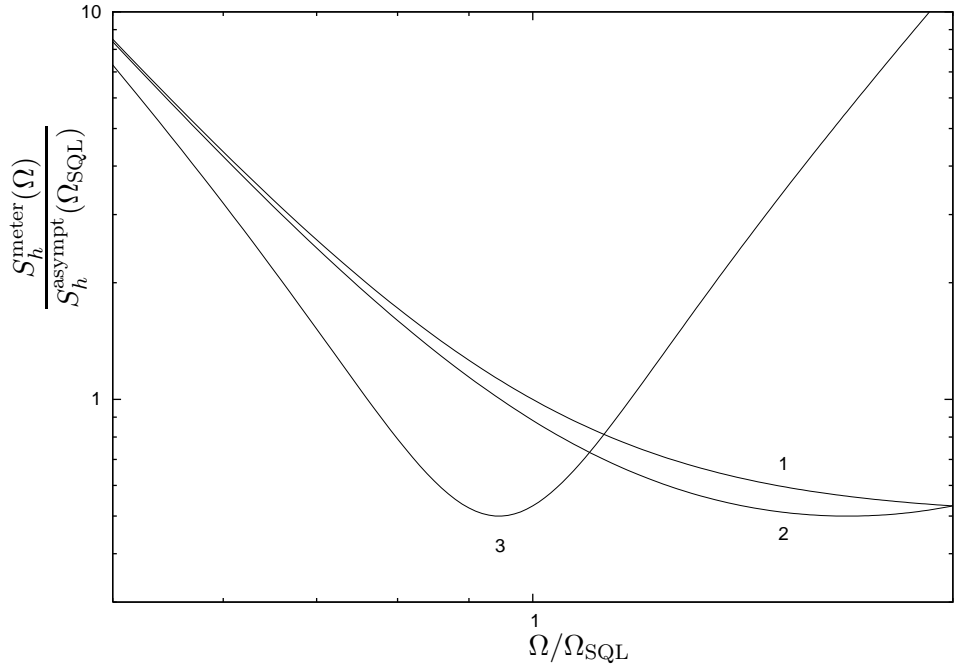}

\end{center}

\caption{Spectral density of the noise of the SQL-limited local meter for
different values of the optical energy ${\cal E}$: 1 --- asymptotic curve for
${\cal E}\to\infty$; 2 --- the optimal energy; 3 --- the energy is too
small.}\label{S_total}

\end{figure}

Suppose that a SQL-limited position meter with frequency-independent
measurement noise $y_{\rm meter}$ and back-action noise $F_{\rm meter}$ are
used as the local meter. For example, a small-scale optical interferometer
with the length $l\ll L$ can be used as such a meter. Suppose also that these
noises are cross-correlated [compare with formula (\ref{correl_simple})]:

\begin{equation}\label{correl}
  \hat y_{\rm meter} = \hat y_{\rm meter}^{(0)}
    - \frac{2M}{2M+m}\,\frac{\hat F_{\rm meter}}{K} \,,
\end{equation}
This simple frequency-independent cross-correlation can be created rather
easily by using a homodine detector with the fixed local oscillator phase
\footnote{In principle, the noises $y^{(0)}_{\rm meter}$ and $F_{\rm meter}$
can also be made cross-correlated by this method. It can be shown, however,
that this additional cross-correlation does not provide any significant
advantages so we do not consider this possibility here.}.

Spectral densities of the measurement noise $S_y$ and the back-action noise
$S_F$ satisfy the uncertainty relation

\begin{equation}
  S_yS_F = \frac{\hbar^2}{4} \,,
\end{equation}
and their ratio depends on the optical pumping energy ${\cal E}_{\rm local}$
in the meter:

\begin{equation}\label{E_local}
  \frac{S_F}{S_y} 
  = \left(\frac{8Q{\cal E}_{\rm local}}{l^2}\right)^2 \,,
\end{equation}
where $Q$ is the quality factor of the local meter cavity.

The spectral density $S_h^{\rm meter}(\Omega)$ of the noise $h_{\rm meter}$
for this meter has rather sophisticated spectral dependence [see formulae
(\ref{S_meter}), (\ref{Spectra})] which allows in principle to obtain
sensitivity better than the SQL in some narrow band. In this paper, however,
wide-band optimization only will be considered.

It is easy to see that if the pumping energy ${\cal E}$ is very large (${\cal
E}\ge {\cal E}_{\rm SQL}$) then the local meter output is close to the output
of the meter attached directly to the test mass $2M+m$ with the signal force
$ML\ddot h$ acting on this mass. Spectral density of the local meter noise
$h_{\rm meter}$ in this case is equal to

\begin{equation}\label{S_meter_SQL}
  S_h^{\rm asympt}(\Omega) = \frac{1}{L^2}
    \left[\frac{S_F}{M^2\Omega^4} + \left(\frac{2M+m}{M}\right)^2S_y\right]
  \ge \frac{2M+m}{M}\,\frac{\hbar}{M\Omega^2L^2} \,.
\end{equation}
(see curve 1 in FIG.\,\ref{S_total}). The rightmost part of this formula
corresponds to the SQL for the considered scheme which, as well as the SQL
(\ref{SQL}), is reached at some specific frequency $\Omega=\Omega_{\rm SQL}$
only. In the case considered here this frequency is equal to

\begin{equation}
  \Omega_{\rm SQL} = \sqrt[4]{\frac{1}{(2M+m)^2}\,\frac{S_F}{S_x}}
  = \sqrt{\frac{8Q{\cal E}_{\rm local}}{(2M+m)l^2}} \,.
\end{equation}
It should be noted that if $m\ll M$ then the SQL (\ref{S_meter_SQL})
corresponds to $\sqrt2$ times better sensitivity than the SQL for traditional
schemes (\ref{SQL}). This gain is obtained because two mirrors $M$ are
required for this scheme instead of four ones as in traditional schemes.

At the same time if ${\cal E}$ is too small then spectral density
(\ref{Spectra_a}) of back-action noise increases sharply at high frequencies;
however, some narrow-band gain in sensitivity can be obtained in this regime
(see curve 3 in FIG.\,\ref{S_total}).

Reasonable intermediate value of ${\cal E}$ can be chosen using, for example,
the following algorithm: (i) require that $S_h^{\rm meter}(\Omega)$ does not
exceeds $S_h^{\rm asympt}(\Omega)$ all over the spectral range of interest, up
to some given frequency $\Omega_{\rm max}$ (see curve 2); (ii) with respect to
this requirement set ${\cal E}$ as small as possible; in any case it have to
be much smaller than ${\cal E}_{\rm SQL}$. In addition condition
(\ref{stability}) which is necessary for the system dynamic stability (see
appendix D of the article \cite{97a1BrGoKh}) is also should be taken into
account.

It is easy to show that these requirement can be satisfied only if
the sloshing frequency is large,

\begin{equation}
  \Omega_B\gg\Omega_{\rm max} \,,
\end{equation}
and in this case the optimal values are equal to

\begin{gather}
  m^* \equiv \frac{2Mm}{2M+m} = 12M\frac{\Omega_{\rm max}^4}{\Omega_B^4} \,,\\
  {\cal E} = \frac{3}{2}\frac{ML^2\Omega_{\rm max}^4}{\omega_o\Omega_B}
  = 3{\cal E}_{\rm SQL}\frac{\Omega_{\rm max}}{\Omega_B} \label{E_SQLmeter} \,.
\end{gather}
Suppose, for example, that $T\approx 0.1$, $L=4\,{\rm Km}$, and therefore
$\Omega_B\approx 7.5\times 10^3\,{\rm s}^{-1}$. In this case if $\Omega_{\rm
max}=2\pi\times 10^2\,{\rm s}^{-1}$, then ${\cal E}\approx 0.25{\cal E}_{\rm
SQL}$ and $m \approx m^*\approx 25\,{\rm g}$.

In principle, more transparent mirror {\sf C} can be used, which allows
further decrease of ${\cal E}$. However, ${\cal E}$ depends on $\Omega_B$ as
$\Omega_B^{-1}$ only, while $m^*$ depends as $\Omega_B^{-4}$. Therefore, the
smaller values of ${\cal E}$ correspond  to very small (sub-gram) values of
the mass $m$. It is unclear, if such a small mirror could tolerate tens of the
kilowatts of the optical power reflecting from it and hundreds of watts
passing through it.

\subsection{QND local meter}

The small-scale optical interferometer discussed in the previous subsection
can be converted into a QND meter by using, for example, so-called
Stroboscopic-Variation Measurement (SVM) technique (see articles \cite{98a1Vy,
00a1DaKhVy, 02a1DaKh}). This method permits to filter out the back-action
noise by using periodic modulation of the local oscillator phase and/or the
pumping energy ${\cal E}_{\rm local}$ with frequency which has to be higher
than the upper frequency of the gravitational-wave signal. For small-scale
interferometers this modulation can be implemented rather easily.

The residual noise spectral density will be proportional to the spectral
density $S_y$ of the meter measurement noise $y_{\rm fluct}$ and in principle
can be made arbitrarily small [see formulae (\ref{Spectra_b},
\ref{QND_meter})] by reducing $S_y$ ({\em i.e.} by increasing the local meter
sensitivity). However, for technological reasons it is useful to provide the
value of the local mirror signal displacement $y_{\rm grav}$ as large as
possible [see formula (\ref{y_grav})].

An optimization algorithm similar to one considered in the previous section
can be used here. Start with the simple quasi-static (low-frequency) case,
when $m^*\Omega^2 \ll K$. In this case the signal displacement is equal to

\begin{equation}\label{y_static}
  y_{\rm grav} = \frac{M}{2M+m}\,Lh \,.
\end{equation}
and corresponding measurement noise is equal to

\begin{equation}
  S_{\rm meas}^{\rm asympt} = \frac{1}{L^2}\left(\frac{2M+m}{M}\right)^2S_y\,.
\end{equation}
Require now that $S_{\rm meas}(\Omega)$ does not exceeds $S_{\rm meas}^{\rm
asympt}$ for all frequencies within the range $0\le\Omega\le\Omega_{\rm max}$.
It is shown in the appendix \ref{QND_opt} that in this case the pumping energy
has to satisfy the following inequality:

\begin{equation}\label{E_QND}
  {\cal E} \ge k\,\frac{m^*L^2\Omega_B\Omega_{\rm max}^2}{2\omega_o}
\end{equation}
where $k$ is a numerical factor which varies from $1/8$ when
$\Omega_B=\Omega_{\rm max}$ to $1/2$ when $\Omega_B\gg\Omega_{\rm max}$ [see
formula (\ref{opt_k})]. This inequality together with the stability condition
(\ref{stability}) can be rewritten as the following condition for the mass
$m^*$:

\begin{equation}\label{m_star}
  \frac{1}{4}\,\biggl(\frac{\Omega_{\rm max}}{\Omega_B}\biggr)^3\,
    \frac{{\cal E}}{{\cal E}_{\rm SQL}}
  \le \frac{m^*}{M} \le
  \frac{1}{k}\,\frac{\Omega_{\rm max}}{\Omega_B}\,
    \frac{{\cal E}}{{\cal E}_{\rm SQL}} \,.
\end{equation}
Values of $m^*$ and $\Omega_B$ in formula (\ref{E_QND}) can vary in wide range
and should be chosen considering technological reasons. In principle, heavy
local mirror {\sf C} with low transmittance $T$ ({\em i.e} low sloshing
frequency) as well as relatively small one with high transmittance $T$ ({\em
i.e} high sloshing frequency) can be used.

Suppose that ${\cal E}=0.1{\cal E}_{\rm SQL}$. This value looks like the
reasonable one due to limitation caused by the internal losses, see subsection
(\ref{opt_losses}). Suppose also that $M=40\,{\rm Kg}$ and $\Omega_{\rm
max}=2\pi\times 100\,{\rm s}^{-1}$. In this case typical numerical examples
are the following.

Heavy local mirror:

\begin{gather}
  T = 0.01 \,, \nonumber \\
  \Omega_B \approx 750\,{\rm s}^{-1} \,, \nonumber \\
  10\,{\rm Kg} \lesssim m^* \lesssim 16\,{\rm Kg} \label{heavyC} \,;
\end{gather}
small local mirror:

\begin{gather}
  T = 0.1 \,, \nonumber \\
  \Omega_B \approx 7500\,{\rm s}^{-1} \,, \nonumber \\
  10\,{\rm g} \lesssim m^* \lesssim 700\,{\rm g} \label{lightC} \,.
\end{gather}
Of course all intermediate values between these two examples are
also possible.

\subsection{The ``optical lever'' scheme}\label{sec:optlever}

In principle, the same idea of the local meter with cross-correlated noise can
be used in order to reduce optical pumping energy in the ``optical lever''
scheme too. However, due to technological limitations in the case of
SQL-limited local meter only modest advantages can be obtained. Really, the
sloshing frequency in the ``optical lever'' scheme is equal to

\begin{equation}
  \Omega_B = \frac{cT}{L\digamma}
  < \frac{7.5\times 10^{4}\,{\rm s}^{-1}}{\digamma}\,.
\end{equation}
(it is supposed that $L=4Km$). At the same time it follows from the formula
(\ref{E_SQLmeter}) that the sloshing frequency have to be equal to

\begin{equation}
  \Omega_B = \frac{3{\cal E}_{\rm SQL}}{{\cal E}}\,\Omega_{\rm max}
  \approx 2\times 10^{4}\,{\rm s}^{-1}\,\frac{{\cal E}_{\rm SQL}}{{\cal E}}
  \,.
\end{equation}
(it is supposed that $\Omega_{\rm max}=2\pi\times 10^2\,{\rm s}^{-1}$). Due to
these limitations it is impossible to obtain significant gain $\digamma$ in
the signal displacement using the pumping energy ${\cal E}<{\cal E}_{\rm
SQL}$.

On the other hand, in the case of a QND local meter low sloshing frequency
$\Omega_B\simeq\Omega_{\rm max}$ can be used [see formulae (\ref{heavyC})]
which makes it possible obtain the gain up to

\begin{equation}
  \digamma = \frac{cT}{L\Omega_{\rm max}} \simeq 10^2 \,.
\end{equation}
The local mirror mass in the case of the ``optical lever'' scheme have to be
$\digamma^2$ time smaller than the figures of formula (\ref{heavyC}). It
follows from the estimates (\ref{heavyC}) that it has to be equal to about
$1\,{\rm g}$. This value does not seems unrealistic one as the optical power
falling on it will be reduced by the factor of $\digamma$ and for the case of
${\cal E}\approx 0.1{\cal E}_{\rm SQL}$ will be equal to about few hundred
watts.

\section{Conclusion}

In the authors opinion, regime of the optical bars/optical lever intracavity
topologies considered in this paper looks rather promising for implementing in
the third generation of gravitational-wave antennae. It allows to obtain
sensitivity better than the SQL and it can do it using rather moderate value
of the optical pumping energy: just tens of kilowatts of circulating power,
instead of megawatts or tens of megawatts.

At the same time, before the implementation of this method, several issues of
a technological origin have to be solved. Some of them are common for all
proposed topologies of the laser gravitational-wave antennae: the problem of
an internal noises of different origin in the test masses (see, for example,
papers \cite{99a1BrGoVy, 00a1BrGoVy, LiuThorne2000, 03a1BrVy}) is the most
notable one, and some are specific for the intracavity topologies. It seems
that the most important of them is the design of the local meter, and it is
evident that this device has to be explored intensively, both experimentally
and theoretically before the decisions about the design of the third
generation laser gravitational-wave antennae will be made.

\section*{Acknowledgments}

Author thanks V.B.Braginsky, S.L.Danilishin, M.L.Gorodetsky and S.P.Vyatchanin
for useful remarks.

This paper was supported in part by NSF and Caltech grant \#PHY0098715, by the
Russian Foundation for Basic Research, and by the Russian Ministry of Industry
and Science.

\appendix

\section{System Dynamics}\label{app:dynamics}

\begin{figure}[t]

\begin{center}

\psfrag{M}[cc][cc]{$M$}
\psfrag{m}[cc][cc]{$m$}
\psfrag{x1}{$x_1$}
\psfrag{x2}{$x_2$}
\psfrag{y}{$y$}
\psfrag{F}[cb]{$\frac12F_{\rm grav}$}

\includegraphics[width=4in]{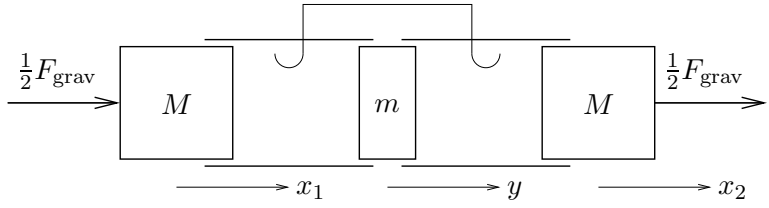}

\end{center}

\caption{Two cavities model}\label{fig:cavities}

\end{figure}

In this appendix we will consider a simplified model of the system similar to
one used in the original paper \cite{97a1BrGoKh}. Replace the Fabry-Perot
cavities {\sf AC} and {\sf BC} (see FIG.\,\ref{fig:optbar}) by two coupled
single-mode e.m. cavities which eigenfrequencies depend on the positions on
the mirrors, see FIG.\,\ref{fig:cavities}

Equations of motion for this system looks like:

{\allowdisplaybreaks\begin{subequations}
  \begin{gather}
    \ddot{\hat q}_1(t) + 2\gamma\dot{\hat q}_1(t)
      + \omega_o^2\left(1+\dfrac{\hat x_1(t)-\hat y(t)}{L}\right)^2\hat q_1(t)
      + \omega_o\Omega_B\hat q_2(t)
      = \dfrac{\omega_o}{\rho}\,[U_{\rm pump}(t) + \hat U_1(t)] \,, \\
    \ddot{\hat q}_2(t) + 2\gamma\dot{\hat q}_2(t)
      + \omega_o^2\left(1-\dfrac{\hat x_2(t)-\hat y(t)}{L}\right)^2\hat q_2(t)
      + \omega_o\Omega_B\hat q_1(t)
      = \dfrac{\omega_o}{\rho}\,[U_{\rm pump}(t) + \hat U_2(t)] \,, \\
    M\ddot{\hat x}_1(t) = -\frac{\omega_o\rho\hat q_1^2(t)}{L}
      + \frac{F_{\rm grav}(t)}{2} \,,\\
    M\ddot{\hat x}_2(t) = \frac{\omega_o\rho\hat q_2^2(t)}{L}
      + \frac{F_{\rm grav}(t)}{2} \,, \\
    m\ddot{\hat y}(t) = \frac{\omega_o\rho}{L}
      \left(\hat q_1^2(t)-\hat q_2^2(t)\right)
      + \hat F_{\rm meter}(t) \,,
  \end{gather}
\end{subequations}}
where $\hat q_{1,2}$ are the generalized coordinates of the cavities, $U_{\rm
pump}$ is the pumping voltage,  $\Omega_B$ is the sloshing frequency which is
proportional to the coupling of the cavities, $\gamma$ is the damping rate of
the cavities, $\hat U_{1,2}$ are the corresponding fluctuational voltages,
$\hat x_{1,2}$ are the positions of the masses $M$, $\hat y$ is the position
of the mass $m$, $F_{\rm grav}$ is the signal force, $\hat F_{\rm meter}$ is
the back-action force of the local meter which monitors variable $y$ (not
shown in the picture).

Let introduce new variables then

\allowdisplaybreaks\begin{subequations}
  \begin{align}
    \hat q_\pm(t) &= \frac{\hat q_1(t) \pm \hat q_2(t)}{\sqrt 2} \,, &
    \hat U_\pm(t) &= \frac{\hat U_1(t) \pm \hat U_2(t)}{\sqrt 2} \,, \\
    \hat x(t) &= \frac{\hat x_1(t)+\hat x_2(t)}{2} \,, &
    \hat X(t) &= \frac{\hat x_1(t)-\hat x_2(t)}{2} \,.
  \end{align}
\end{subequations}
For these variables we obtain:

{\allowdisplaybreaks\begin{subequations}\label{A3}
  \begin{gather}
    \ddot{\hat q}_+(t) + 2\gamma\dot{\hat q}_+(t) + \omega_+^2\hat q_+(t)
      + \frac{2\omega_o^2}{L}
          \Bigl(\hat X(t)\hat q_+(t) + [\hat x(t)-\hat y(t)]\hat q_-(t)\Bigr)
      \nonumber \\ \qquad
      = \dfrac{\omega_o}{\rho}\,[\sqrt2\,U_{\rm pump}(t) + \hat U_+(t)] \,, \\
    \ddot{\hat q}_-(t) + 2\gamma\dot{\hat q}_-(t) + \omega_-^2\hat q_-(t)
      + \frac{2\omega_o^2}{L}
          \Bigl([\hat x(t)-\hat y(t)]\hat q_+(t) + \hat X(t)\hat q_-(t)\Bigr)
      = \dfrac{\omega_o}{\rho}\,U_-(t) \,, \\
    2M\ddot{\hat x}(t) = -\frac{2\omega_o\rho}{L}\,\hat q_+(t)\hat q_-(t)
      + F_{\rm grav}(t) \,,\\
    2M\ddot{\hat X}(t)
      = -\frac{\omega_o\rho}{L}[\hat q_+^2(t)+\hat q_-^2(t)] \,, \\
    m\ddot{\hat y}(t) = \frac{2\omega_o\rho}{L}\,\hat q_+(t)\hat q_-(t)
      + \hat F_{\rm meter}(t) \,,
  \end{gather}
\end{subequations}}
where $\omega_\pm = \omega_o\pm \Omega_B/2$.

Suppose that the pumping frequency is equal to $\omega_+$ and amplitude of the
pumping field in the mode ``+'' is equal to $q_0$. Keeping only linear in
$q_0$ term in the right parts of the equations (\ref{A3}), these equation can
be rewritten as:

{\allowdisplaybreaks\begin{subequations}
  \begin{gather}
    q_+(t) = q_0\cos\omega_+t \,, \\
    \ddot{\hat q}_-(t) + 2\gamma\dot{\hat q}_-(t) + \omega_-^2\hat q_-(t)
      = \dfrac{\omega_o}{\rho}\,\hat U_-(t)
        + \dfrac{2\omega_o^2q_0}{L}\,[\hat y(t)-\hat x(t)]\cos\omega_+t \,, \\
    2M\ddot{\hat x}(t) = -\frac{2\omega_o\rho q_0}{L}\,\hat q_-(t)\cos\omega_+t
      + F_{\rm grav}(t) \,,\\
    m\ddot{\hat y}(t) = \frac{2\omega_o\rho q_0}{L}\,\hat q_-(t)\cos\omega_+t
      + \hat F_{\rm meter}(t)\,.
  \end{gather}
\end{subequations}}
Using then the rotation polarization approximation:

{\allowdisplaybreaks\begin{subequations}
  \begin{gather}
    \hat q_-(t) = \hat q_c(t)\cos\omega_+t + \hat q_s(t)\sin\omega_+t \,, \\
    \dot{\hat q}_-(t)
      = \omega+(\hat q_c(t)\cos\omega_+t + \hat q_s(t)\sin\omega_+t) \,, \\
    \hat U_-(t) = \hat U_c(t)\cos\omega_+t + \hat U_s(t)\sin\omega_+t \,,
  \end{gather}
\end{subequations}}
we obtain a simple linear equations set which is convenient for
spectral representation:

{\allowdisplaybreaks\begin{subequations}\label{GB_114}
  \begin{gather}
    (i\Omega + \gamma)\hat q_c(\Omega) + \Omega_B\hat q_s(\Omega)
      = - \dfrac{\hat U_s(\Omega)}{2\rho} \,, \\
    - \Omega_B\hat q_c(\Omega) + (i\Omega + \gamma)\hat q_s(\Omega)
      = \dfrac{\hat U_c(\Omega)}{2\rho}
        + \frac{\omega_oq_0}{L}[\hat y(\Omega)-\hat x(\Omega)]\,,\\
    -2M\Omega^2\hat x(\Omega) = -\frac{\omega_o\rho q_0}{L}\,\hat q_c(t)
      + F_{\rm grav}(\Omega) \,,\\
    -m\Omega^2\hat y(\Omega) = \frac{\omega_o\rho q_0}{L}\,\hat q_c(\Omega)
      + \hat F_{\rm meter}(\Omega) \,.
  \end{gather}
\end{subequations}}
From the first two equations we obtain:

\begin{equation}
  \hat q_c(\Omega) = -\frac{1}{{\cal D}(\Omega)}\biggl(
    \frac{(i\Omega+\gamma)\hat U_s(\Omega)+\Omega_B\hat U_c(\Omega)}{2\rho}
    + \dfrac{\omega_o\Omega_Bq_0}{L}[\hat y(\Omega)-\hat x(\Omega)]
  \biggr) \,,
\end{equation}
where

\begin{equation}
  {\cal D}(\Omega) = (i\Omega+\gamma)^2+\Omega_B^2 \,.
\end{equation}
Substitution of this value of $q_c$ into the last two equations of
(\ref{GB_114}) gives:

\begin{subequations}\label{mech_eq}
  \begin{gather}
    [-2M\Omega^2 + {\cal K}(\Omega)]\hat x(\Omega)
      - {\cal K}(\Omega)\hat y(\Omega)
      = \hat F_{\rm loss}(\Omega) + F_{\rm grav}(\Omega) \,, \\
    [-m\Omega^2 + {\cal K}(\Omega)]\hat y(\Omega)
      - {\cal K}(\Omega)\hat x(\Omega)
      = -\hat F_{\rm loss}(\Omega) + \hat F_{\rm meter}(\Omega) \,,
  \end{gather}
\end{subequations}
where

\begin{equation}\label{cal_K}
  {\cal K}(\Omega) = \frac{2\omega_o{\cal E}\Omega_B}{L^2{\cal D}(\Omega)}
  = \frac{K\Omega_B^2}{{\cal D}(\Omega)}
\end{equation}
is the complex pondermotive rigidity,

\begin{equation}
  {\cal E} = \frac{\omega_o\rho q_0^2}{2}
\end{equation}
is the pumping energy and

\begin{equation}\label{F_loss}
  \hat F_{\rm loss}(\Omega) = \frac{\omega_o q_0}{2L{\cal D}(\Omega)}
    \Bigl((i\Omega+\gamma)U_s(\Omega)+\Omega_BU_c(\Omega)\Bigr)
\end{equation}
is the fluctuational force which arises due to losses in the cavities.
Spectral densities of $\hat U_{c,s}(t)$ are equal to

\begin{equation}
  S_{U_{c,s}} = 4\hbar\rho\gamma \,,
\end{equation}
therefore, spectral density of $F_{\rm loss}(t)$ is equal to

\begin{equation}\label{S_loss_raw}
  S_{F\,\rm loss} = \frac{2\hbar\omega_o{\cal E}\gamma}{L^2}\,
    \frac{\Omega^2+\gamma^2+\Omega_B^2}{\lvert{\cal D}(\Omega)\rvert^2} \,.
\end{equation}
It follows from the equations (\ref{mech_eq}), that the position of the local
mass is equal to

\begin{equation}\label{tilde_y_2}
  \hat y(\Omega) = y_{\rm grav}(\Omega) +
  \frac{
    [{\cal K}(\Omega)-2M\Omega^2]\hat F_{\rm meter}(\Omega)
    + 2M\Omega^2\hat F_{\rm loss}(\Omega)
  }{-(2M+m)\Omega^2[{\cal K}(\Omega)-m^*\Omega^2]} \,,
\end{equation}
where

\begin{equation}\label{y_grav_raw}
  y_{\rm grav}(\Omega) = \frac{M}{2M+m}\,
    \frac{{\cal K}(\Omega)}{{\cal K}(\Omega)-m^*\Omega^2}\,Lh(\Omega)
\end{equation}
is the signal displacement and

\begin{equation} m^* = \frac{2Mm}{2M+m} \,. \end{equation}

It have to be noted also that as it was shown in the article \cite{97a1BrGoKh}
this system is dynamically unstable. If

\begin{equation}\label{stability}
  K \le \frac{m^*\Omega_B^2}{4} \,,
\end{equation}
then this instability is relatively small and can be rather easily suppressed
by a feed-back system. In the opposite case, however, very strong asynchronous
instability arises (see appendix D of the article \cite{97a1BrGoKh}) which
makes the scheme virtually useless. Therefore, condition (\ref{stability}) has
to be considered as necessary one.

\section{The output signal}\label{app:noises}

The output signal of the local meter is equal to

\begin{equation}
  \tilde y(\Omega) = \hat y(\Omega) + \hat y_{\rm meter}(\Omega) \,,
\end{equation}
where $\hat y_{\rm meter}(\Omega)$ is the measurement noise.

If the meter noises are cross-correlated, see formula (\ref{correl}), then

\begin{multline}
  \tilde y(\Omega) = y_{\rm grav}(\Omega) +
  \frac{1}{-(2M+m)\Omega^2[{\cal K}(\Omega)-m^*\Omega^2]} \\
  \times\biggl\{
    \left[{\cal K}(\Omega)
      + 2M\Omega^2\left(\frac{{\cal K}(\Omega)}{K}-1\right)
      - \frac{2Mm^*\Omega^4}{K}
      \right]\hat F_{\rm meter}(\Omega)
    + 2M\Omega^2\hat F_{\rm loss}(\Omega)
  \biggr\} + \hat y_{\rm meter}^{(0)}(\Omega) \,,
\end{multline}
It is convenient to present this expression as follows:

\begin{equation}
  \tilde y(\Omega)
  = \frac{M}{2M+m}\,\frac{{\cal K}(\Omega)L}{{\cal K}(\Omega)-m^*\Omega^2}
    \left[
      h(\Omega) + \hat h_{\rm loss}(\Omega) + \hat h_{\rm meter}(\Omega)
    \right] \,,
\end{equation}
where

\begin{equation}\label{h_loss}
  \hat h_{\rm loss}(\Omega)
    = - \frac{2F_{\rm loss}(\Omega)}{{\cal K}(\Omega)L}
\end{equation}
is the equivalent noise produced by the optical losses, and

\begin{multline}\label{h_meter_raw}
  \hat h_{\rm meter}(\Omega) = \frac{1}{L}\biggl\{
    \left[
      -\frac{1}{M\Omega^2}
      - 2\left(\frac{1}{K}-\frac{1}{{\cal K}(\Omega)}\right)
      + \frac{2m^*\Omega^2}{K{\cal K}(\Omega)}
    \right]\hat F_{\rm meter}(\Omega) \\
    + \frac{2M+m}{M}\left(1-\frac{2m^*\Omega^2}{{\cal K}(\Omega)}\right)
        \hat y_{\rm meter}(\Omega)
  \biggr\}
\end{multline}
is the equivalent noise of the meter (both these noises are normalized as an
equivalent fluctuational gravitational-wave signals).

Taking into account that in order to obtain $\xi_{\rm loss}^2<1$ [see formula
(\ref{xi2_loss})] the optical losses have to be small, $\gamma\ll\Omega$,
expressions (\ref{y_grav_raw}), (\ref{h_meter_raw}) can be slightly
simplified:

\begin{equation}\label{y_grav}
  y_{\rm grav}(\Omega) = \frac{M}{2M+m}\,
    \frac{Lh(\Omega)}
      {1-\dfrac{m^*\Omega^2(\Omega_B^2-\Omega^2)}{K\Omega_B^2}}\, \,,
\end{equation}
\begin{multline}\label{OBG:h_meter}
  \hat h_{\rm meter}(\Omega) = \frac{1}{L}\biggl\{
    \left[
      -\frac{1}{M\Omega^2} - \frac{2\Omega^2}{K\Omega_B^2}
      + \frac{2m^*\Omega^2(\Omega_B^2-\Omega^2)}{K^2\Omega_B^2}
    \right]\hat F_{\rm meter}(\Omega) \\
    + \frac{2M+m}{M}\left[
        1 - \frac{m^*\Omega^2(\Omega_B^2-\Omega^2)}{K\Omega_B^2}
      \right]\hat y_{\rm meter}^{(0)}(\Omega)
  \biggr\}
\end{multline}

If a SQL-limited local meter is used, then spectral density of
this noise can be presented as a sum

\begin{equation}\label{S_meter}
  S_h^{\rm meter}(\Omega) = S_{\rm B.A.}(\Omega) + S_{\rm meas}(\Omega) \,,
\end{equation}
where

\begin{subequations}\label{Spectra}
  \begin{gather}
    S_{\rm B.A.}(\Omega) = \frac{1}{L^2}\left[
        -\frac{1}{M\Omega^2} - \frac{2\Omega^2}{K\Omega_B^2}
        + \frac{2m^*\Omega^2(\Omega_B^2-\Omega^2)}{K^2\Omega_B^2}
      \right]^2S_F \,, \label{Spectra_a}\\
    S_{\rm meas}(\Omega) = \frac{1}{L^2}\left(\frac{2M+m}{M}\right)^2\left[
        1 - \frac{m^*\Omega^2(\Omega_B^2-\Omega^2)}{K\Omega_B^2}
      \right]^2S_y \label{Spectra_b}
 \end{gather}
\end{subequations}
and $S_F$ and $S_y$ are spectral densities of the meter back-action noise
$\hat F_{\rm meter}$ and its measurement noise $\hat y_{\rm meter}(\Omega)$.

\section{Noise optimization for a QND local meter}\label{QND_opt}

In the case of a QND local meter the back-action noise can be filtered out by
some means and sensitivity is limited by the measurement noise only:

\begin{equation}\label{QND_meter}
  S_h^{\rm meter}(\Omega) = S_{\rm meas}(\Omega) \,.
\end{equation}
Require that $S_{\rm meas}(\Omega)$ has not to exceed $S_{\rm meas}(0)$ for
all frequencies $0\le\Omega\le\Omega_{\rm max}$. In this case the following
condition has to be fulfilled for all these frequencies:

\begin{equation}
  \left|1 - \frac{m^*\Omega^2(\Omega_B^2-\Omega^2)}{K\Omega_B^2}\right| \le 1
  \,.
\end{equation}
The solution of this inequality can be presented as follows:

\begin{equation}
  K \ge km^*\Omega_{\rm max}^2 \,,
\end{equation}
where

\begin{equation}\label{opt_k}
  k = \begin{cases}
    \dfrac{1}{8}\biggl(\dfrac{\Omega_B}{\Omega_{\rm max}}\biggr)^2 \,,
      & \Omega_{\rm max} \le \Omega_B \le \sqrt2\Omega_{\rm max} \medskip \\
    \dfrac{1}{2}
      \biggl[1-\biggl(\dfrac{\Omega_{\rm max}}{\Omega_B}\biggr)^2\biggr] \,,
      & \Omega_B > \sqrt2\Omega_{\rm max} \,.
  \end{cases}
\end{equation}



\end{document}